\PassOptionsToPackage{dvipsnames,table}{xcolor}
\PassOptionsToPackage{pdfa}{hyperref}

\RequirePackage[bookmarksnumbered,unicode]{hyperref}
\RequirePackage{hyperxmp}

\documentclass[sigconf]{acmart}
\hypersetup{pdfapart=2, pdfaconformance=u}
\AtBeginDocument{%
  }

\usepackage{tikz}
\newcommand{\copyrighttext}{ \scriptsize \textcopyright 2023 ACM.               
	Personal use of this material is permitted.                                 
	Permission from ACM must be obtained for all other uses,                   
	in any current or future media, including reprinting/republishing this      
	material for advertising or promotional purposes, creating new collective   
	works, for resale or redistribution to servers or                           
	lists, or reuse of any copyrighted component of this work in other works.   
	This is the author’s version of the work. The final version is published in the proceedings of the 39th {ACM/SIGAPP} Symposium on Applied Computing (SAC). DOI: \href{https://doi.org/10.1145/3605098.3635994}{10.1145/3605098.3635994}}

\setcopyright{none}
\setcopyright{none}
\renewcommand\footnotetextcopyrightpermission[1]{}

\usepackage[utf8]{inputenc}
\usepackage[T1]{fontenc}
\usepackage{algorithm}
\usepackage{algpseudocode}
\usepackage{graphicx}
\usepackage[margin=3pt,skip=3pt,font={small,stretch=0.9},labelfont=bf]{caption}
\usepackage{xcolor}

\usepackage{natbib}
\usepackage{multirow}

\usepackage{amsmath}
\usepackage{amsfonts}
\usepackage{setspace}
\usepackage{pifont} %\ding
\usepackage[abbreviations]{foreign}
\usepackage{tabularx}

\usepackage{textcomp}
\usepackage{booktabs}
\usepackage{xspace}
\usepackage{url}

\usepackage[inline]{enumitem}

\usepackage[capitalise]{cleveref}

\usepackage{multicol}
\definecolor{lightcolor}{rgb}{0,0.5,1}
\usepackage{tcolorbox}
\usepackage{adjustbox}

\newcounter{numobserv} 
\definecolor{beaublue}{rgb}{0.88, 0.93, 0.93}
\usepackage{tcolorbox}
\colorlet{shadecolor}{beaublue}
\tcbset{
	colback=beaublue,
	boxrule=0.5pt,
	boxsep=0pt,
	left=1pt,
	right=1pt,
	top=1pt,
	bottom=1pt,
	sharp corners,
	colframe=white,
}
\newcommand{\observ}[1]{
	\addtocounter{numobserv}{1}
	\begin{tcolorbox}	
		\textit{\textbf{Take-away\,\thenumobserv\,:} #1 }	
	\end{tcolorbox}
}

\usepackage{fontawesome5}
\newboolean{showcomments}
\setboolean{showcomments}{true}
\ifthenelse{\boolean{showcomments}}
{ \newcommand{\mynote}[3]{
		\fbox{\bfseries\sffamily\scriptsize#1}
		{\small$\blacktriangleright$\textsf{\emph{\color{#3}{#2}}}$\blacktriangleleft$}}}
{ \newcommand{\mynote}[3]{}}

\definecolor{darkgreen}{rgb}{0.3,0.5,0.3}
\definecolor{darkblue}{rgb}{0.3,0.3,0.5}
\definecolor{darkred}{rgb}{0.5,0.3,0.3}

\definecolor{ggreen}{RGB}{0,100,0}
\definecolor{rred}{RGB}{128,0,32}

\newcommand{\code}[1]{\texttt{\small #1}\xspace}

\newcommand{\subpoint}[1]{\smallskip\noindent\textit{#1}\xspace}
\newcommand{\sys}{\textsc{Fortress}\xspace}
\newcommand{\iis}{\texttt{I$^2$S}\xspace}

\newcommand{\yes}{\textcolor{ggreen}{\ding{51}}}
\newcommand{\no}{\textcolor{rred}{\ding{55}}}

\newcommand{\optee}{\textsc{OP-TEE}\xspace}

\usepackage{listings}
\lstdefinestyle{CStyle}{
	backgroundcolor=\color{backgroundColour},   
	commentstyle=\color{mGreen},
	keywordstyle=\color{magenta},
	numberstyle=\tiny\color{mGray},
	stringstyle=\color{mPurple},
	basicstyle=\footnotesize,
	breakatwhitespace=false,         
	breaklines=true,                 
	captionpos=b,                    
	keepspaces=true,                 
	numbers=left,                    
	numbersep=5pt,                  
	showspaces=false,                
	showstringspaces=false,
	showtabs=false,                  
	tabsize=2,
	language=C
}

\begin{document}

%\title{Fortress: Securing IoT Peripherals with Trusted Execution Environments and Machine Learning}
\title[Fortress: Securing IoT Peripherals with Trusted Execution Environments]{Fortress: Securing IoT Peripherals\\with Trusted Execution Environments}
%\normalsize{Practical Experience Report --- Submission \#311}}

\author{Peterson Yuhala}
\orcid{0000-0002-3371-9228}
\affiliation{%
	\institution{University of Neuchâtel}
	\city{Neuchâtel}
	\country{Switzerland}}
\email{peterson.yuhala@unine.ch}

\author{Jämes Ménétrey}
\orcid{0000-0003-2470-2827}
\affiliation{%
	\institution{University of Neuchâtel}
	\city{Neuchâtel}
	\country{Switzerland}}
\email{james.menetrey@unine.ch}

\author{Pascal Felber}
\orcid{0000-0003-1574-6721}
\affiliation{%
	\institution{University of Neuchâtel}
	\city{Neuchâtel}
	\country{Switzerland}}
\email{pascal.felber@unine.ch}

\author{Marcelo Pasin}
\orcid{0000-0002-3064-5315}
\affiliation{%
	\institution{University of Neuchâtel}
	\city{Neuchâtel}
	\country{Switzerland}}
\email{marcelo.pasin@unine.ch}

\author{Valerio Schiavoni}
\orcid{0000-0003-1493-6603}
\affiliation{%
	\institution{University of Neuchâtel}
	\city{Neuchâtel}
	\country{Switzerland}}
\email{valerio.schiavoni@unine.ch}

\renewcommand{\shortauthors}{Yuhala et al.}

\newcommand{\copyrightnotice}{\begin{tikzpicture}[remember picture,overlay]       
	\node[anchor=south,yshift=2pt,fill=yellow!20] at (current page.south) {\fbox{\parbox{\dimexpr\textwidth-\fboxsep-\fboxrule\relax}{\copyrighttext}}};
	\end{tikzpicture}
}

%!TEX root = main.tex
\begin{abstract} 
With the increasing popularity of Internet of Things (IoT) devices, securing sensitive user data has emerged as a major challenge.
These devices often collect confidential information, such as audio and visual data, through peripheral inputs like microphones and cameras. 
Such sensitive information is then exposed to potential threats, either from malicious software with high-level access rights or transmitted (sometimes inadvertently) to untrusted cloud services.
In this paper, we propose a generic design to enhance the privacy in IoT-based systems by isolating peripheral I/O memory regions in a secure kernel space of a trusted execution environment (TEE).
Only a minimal set of peripheral driver code, resident within the secure kernel, can access this protected memory area.

This design effectively restricts any unauthorised access by system software, including the operating system and hypervisor. 
The sensitive peripheral data is then securely transferred to a user-space TEE, where obfuscation mechanisms can be applied before it is relayed to third parties, \eg, the cloud.
To validate our architectural approach, we provide a proof-of-concept implementation of our design by securing an audio peripheral based on inter-IC sound (\iis), a serial bus to interconnect audio devices.
The experimental results show that our design offers a robust security solution with an acceptable computational overhead.
%, \ie \vs{tease the numbers here}.
\end{abstract}





\maketitle
\copyrightnotice
\pagestyle{plain}

%!TEX root = main.tex
\section{Introduction}
\label{sec:intro}
%\pf{Pascal comment.}
%\vs{Valerio comment.}
%\jm{James comment.}
%\mpas{Marcelo comment.}
%\py{Peterson comment.}

The Internet of Things (IoT)~\cite{iot-luigi} concept has gained tremendous popularity in recent years. In broad terms, IoT refers to the network of smart devices capable of collecting, processing, and exchanging data over the internet. IoT has vast applications across various domains, including smart homes, healthcare, agriculture, industrial automation, and enables a wide range of possibilities, such as remotely controlling devices, monitoring environmental conditions, improving efficiency and enhancing safety.

The widespread adoption of IoT platforms has raised serious security and privacy concerns due to the nature of interconnected devices and the vast amounts of data generated~\cite{apthorpe2017smart}. 
For instance, in smart homes, large amounts of sensitive data are constantly generated through sensors and hardware peripheral devices, \eg cameras, microphones. 
This data is usually transmitted to untrusted cloud services and third-party providers, oftentimes with little regard to the confidentiality of the shared data. 
The involvement of multiple parties introduces privacy issues, as the data could be used for purposes beyond the user's knowledge or control. 
For example, in July 2019, more than 1000 Google Assistant recordings were involuntarily leaked~\cite{google-leak,cnbc}, with part of these recordings activated accidentally by users. Furthermore, privileged software like the underlying operating system (OS) or hypervisor can be compromised~\cite{keystone,VanNostrand2019ConfidentialDL}, potentially leading to data breaches or unauthorized access to sensitive data. 

To cope with these security threats, hardware-based security technologies that allow the creation of \textit{trusted execution environments} (TEEs) have been developed. 
Popular and emerging implementations include Intel software guard extensions (SGX)~\cite{vcostan}, Intel trust domain extensions (TDX)~\cite{cheng2023intel}, and AMD secure encrypted virtualization (SEV)~\cite{sev-whitepaper} which target server-end environments, and ARM TrustZone~\cite{pinto19} or RISC-V MultiZone~\cite{garlati2021secure}, respectively for low-power ARM-based or RISC-V-based client-end devices.
%\vs{it's not clear why user and kernel spaces mentioned next are relevant}
Although TEE frameworks like \optee~\cite{optee} have been leveraged to secure both user and kernel level applications, a holistic and generic solution for safeguarding peripheral data in IoT setups remains absent.
%\vs{you should clarify that the framework offers such user+kernel space shield, if that's the case}
%Prior works have leveraged TEEs to enhance security of peripheral data but have limitations which we aim to address:~\cite{secloak} uses TrustZone technology to reliably turn on/off peripherals on mobile devices, while \cite{driverguard, overshadow, sgxio} leverage the hypervisor to isolate applications and their data from an untrusted kernel. %While these systems provide good security guarantees for their described threat models, they lack the capability to implement precise filtering mechanisms for sensitive data in an IoT setup, preventing unauthorized access by a compromised kernel and hypervisor, or exposure to an untrusted cloud provider.

Our work aims to fill this gap by proposing \sys, a robust and comprehensive framework to enhance security and privacy in IoT infrastructures.
%\vs{devices/Peripherals (as in the title)?}. \py{it is specified in the next phrase.}
Our approach involves restricting peripheral I/O memory to a secure kernel space TEE, providing access only to a small part of peripheral driver code while isolating peripheral data from an untrusted OS or hypervisor. 
The data is then securely transferred to a user space TEE, where obfuscation or data sanitization techniques can be applied to encrypt or remove sensitive information before it is transmitted to an untrusted cloud environment. As we will see in \S\ref{sec:discussion}, this security framework has practical applications in various contexts, including smart homes, medical IoT, retail IoT, amongst many others.

In summary, the contributions of this work are as follows:
\begin{itemize}
	\item A design to secure IoT peripheral data using ARM TrustZone.
	\item A generic approach to partition peripheral drivers for ARM TrustZone.
	\item A proof-of-concept implementation of our design using \iis-based peripherals on ARMv8-A, demonstrating the feasibility of the proposed approach.
	\item A comprehensive evaluation of our system, which shows that privacy is achieved at a reasonable cost.
\end{itemize}

The paper is structured as follows:
\S\ref{sec:background} provides background concepts that underpin our work;
\S\ref{sec:tmodel} describes the threat model;
\S\ref{sec:arch} elaborates on the system architecture of \sys; 
\S\ref{sec:implem} highlights key implementation details;
\S\ref{sec:eval} presents our experimental evaluation;
\S\ref{sec:rw} discusses related work;
\S\ref{sec:discussion} provides practical applications of \sys;
\S\ref{sec:limitations} addresses the limitations of our approach;
and \S\ref{sec:conclusion} offers concluding remarks and future directions.

\section{Background}
\label{sec:background}
This section describes key concepts that underpin our work: \textit{trusted execution environment} (TEE) (\S\ref{ssec:tee}), OP-TEE (\S\ref{ssec:optee}), \textit{memory-mapped I/O} (MMIO) and \textit{direct memory access} (DMA) (\S\ref{ssec:mmio}).

\subsection{Trusted Execution Environments}\label{ssec:tee}

\begin{figure}[!t]
	\centering
	\includegraphics[scale=0.55]{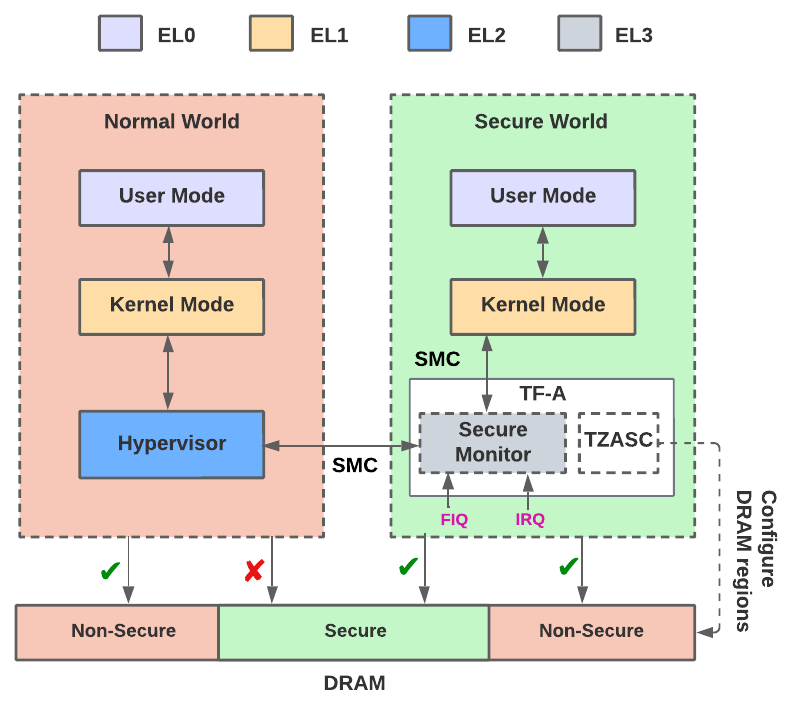}
	\caption{TrustZone security architecture on Arm Cortex-A.}
	\label{fig:tz-arch}
	\vspace{-4mm}
\end{figure}

A \emph{trusted execution environment} (TEE) is a hardware-enforced security technique to isolate sensitive code and data at runtime from potentially malicious entities, including privileged software like the OS or hypervisor. Various hardware security technologies have been developed that allow to create a TEE.

Intel SGX~\cite{vcostan} provides process isolation through \textit{enclaves}, while
AMD SEV~\cite{sev-whitepaper} and more recently Intel TDX~\cite{cheng2023intel} focus on isolating virtual machines (VMs).
These technologies are specifically tailored for server-grade environments.

%Intel SGX~\cite{vcostan} provides a set of x86 instructions that enable applications to create secure encrypted memory regions, called \textit{enclaves}, that can only be read inside the CPU, providing strong security guarantees to applications being executed on remote untrusted servers. 
%Intel TDX~\cite{cheng2023intel} is a new architectural extension in the fourth generation Intel Xeon processor that allows the deployment of hardware-isolated virtual machines (VMs) called \textit{trust domains} with encrypted and integrity-protected memory and CPU state.
%AMD SEV~\cite{sev-whitepaper} also leverages processor instructions to provide strong isolation guarantees to colocated guest VMs in a cloud environment. 
%Intel SGX, Intel TDX, and AMD SEV are adapted for server-grade environments. 

For edge-based and low-power devices, ARM TrustZone (TZ)~\cite{pinto19} is the most common TEE technology. 
TZ is a security technology for ARM-based processors which divides the processor into two protection domains: \textit{secure world} wherein sensitive operations can be performed, and \textit{normal world} for performing non-sensitive operations. 
\autoref{fig:tz-arch} describes TrustZone's security architecture. 
At any point in time, the processor operates exclusively in one of these worlds. 
A special bit known as the \textit{non-secure (NS)} bit stored in the \textit{secure configuration register (SCR)} determines the current protection domain of the processor, and is used for memory access control checks across both worlds. 
The processor can equally transition between worlds; TrustZone introduces a new component known as the \textit{secure monitor} (operating in \textit{monitor mode}) which acts as a bridge between both worlds and is responsible for storing processor state during transitions. 
A new privileged instruction, \ie \textit{secure monitor call (SMC)}, allows software in both worlds to switch to the opposite world via monitor mode. 
Regarding interrupts, IRQ (normal interrupt request) and FIQ (fast interrupt request) in the secure world can also trigger a transition to monitor mode without an SMC. 
The ARMv8 architecture provides four privilege levels, \ie \textit{exception levels} (EL)~\cite{arm-el} at which code can run: EL0 for user space code, EL1 kernel space code, \eg the OS, EL2 for the hypervisor, and EL3 for secure monitor mode. 
\footnote{A prefix of "S" is usually added to the exception level for more precision when the code is being executed in the secure world. For example S-EL0/S-EL1 explicitly indicate that the user space/kernel space code is executing in the secure world.}
%\vs{I would move this into a footnote: A prefix of "S" is usually added to the exception level for more precision when the code is being executed in the secure world. 
%For example S-EL0/S-EL1 explicitly indicate that the user space/kernel space code is executing in the secure world.}

The memory infrastructure in TrustZone-enabled systems (Cortex-A) introduces a hardware component called the \textit{TrustZone address space controller} (TZASC), which the ARM Trusted Firmware (TF-A)~\cite{atf} uses to configure specific DRAM areas as secure regions. 
These configurations can be done such that secure world applications can access all memory regions while normal world applications are confined to non-secure memory. 
A similar memory partitioning functionality is performed by a component called \textit{TrustZone memory adapter} (TZMA) but targets SRAM rather than DRAM. 
Both TZASC and TZMA may or may not exist on a specific system-on-chip (SoC) implementation. 
For example, the Raspberry Pi 3 plaftform supports some ARM TrustZone features but lacks TZASC and TZMA~\cite{tz-mem-attack}, and hence it lacks the capability of securing memory with TrustZone.  

\vspace{-2mm}
\subsection{Open Portable TEE (OP-TEE)}
\label{ssec:optee}
\begin{figure}[!t]
	%\hspace{-1mm}
	\centering
	\includegraphics[scale=0.43]{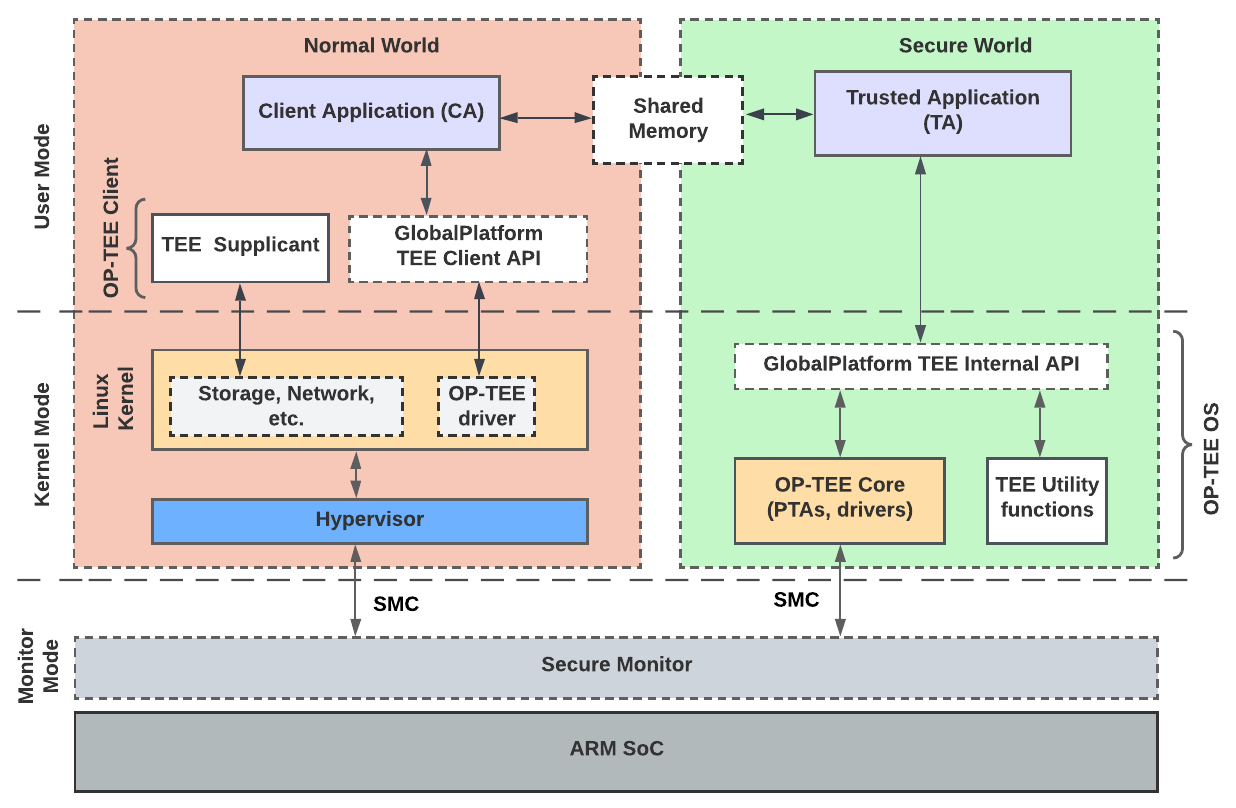}
	\caption{OP-TEE architecture.}
	\label{fig:optee-arch}
	\vspace{-4mm}
	%\vspace{mm}
\end{figure}

OP-TEE~\cite{optee} is a software framework that implements a trusted execution environment for TrustZone.
OP-TEE is compliant with the GlobalPlatform TEE Internal/Client API specification v1.0~\cite{gpf}, and consists of three main components: \textit{OP-TEE client}, \textit{OP-TEE OS}, and \textit{OP-TEE Linux driver}.
OP-TEE client offers an API that allows applications in the normal world, also known as \textit{client application} (CAs), to interact with applications running in the secure world, known as \textit{trusted application} (TAs).
Specifically, the CAs operate within a \textit{rich execution environment} (REE), \ie, in the regular OS at the EL0 level, while TAs execute in the secure world inside the TEE, at the S-EL0 level.
%The secure world comprises ARM Trusted Firmware (TF-A), which contains the secure monitor for handling context switches between both worlds based on SMC exceptions or interrupt notifications. 
%TF-A also handles TZASC configurations at boot time.
OP-TEE Linux driver is a kernel-space component operating in the normal world OS (at EL1) which facilitates transitions between normal and secure worlds. 
In the presence of a hypervisor, the SMC from a guest kernel traps into the hypervisor and the latter performs the SMC on behalf of the former.
OP-TEE OS provides an interface called a \textit{pseudo trusted application} (PTA) which TAs can use to communicate with OP-TEE OS.
 
Given the large dominance of ARM-based architectures on user-end devices in IoT setups, the architecture and design of \sys is based on ARM Trustzone and OP-TEE. % because ARM is the dominant architecture on user-end devices in IoT platforms.

\subsection{MMIO and DMA}\label{ssec:mmio}
%\py{bg on device drivers, device trees, memory mapped io, i2c, i2s, etc. Give example device tree showing base addresses and sizes. Talk about reading device tree files to obtain base addresses vs hardcoding device tree configurations, etc. Check DTS documentation on OPTEE. Hardcoding prevents reading dts files at runtime, but is less flexible.}
%\py{what are peripheral drivers ?}
%\begin{figure}[!t]
%	\centering
%	\includegraphics[scale=0.55]{fig/device-drivers}
%	\caption{Device drivers.}
%	\label{fig:dev-drivers}
%\end{figure}
%\py{basic but useful way to begin the paragraph.}
Contemporary computer systems and IoT devices comprise peripheral devices, \eg keyboard, mouse, camera, microphone, fingerprint sensors, aimed to provide I/O capabilities. Dedicated system software called \textit{device drivers} enable the OS kernel to interact with these peripherals.
The job of a typical device driver is, for the most part, reading or writing I/O memory~\cite{corbet2005linux}.
This is commonly achieved through mechanisms like MMIO and DMA, which we detail next. 

\subpoint{MMIO.~} It is a method of performing I/O between the CPU and a peripheral device by mapping the peripheral's \textit{I/O registers} into the CPU's address space.
As such, processor \code{load}/\code{store} instructions on mapped addresses translate to load/store operations on the corresponding I/O registers in physical memory. 
The Linux kernel on 64-bit ARM platforms provides MMIO access primitives like \code{ioreadX}/\code{iowriteX} which read and write \code{X} bits of data from and to an I/O register, respectively. 
For example, \code{ioread8(reg)} and \code{iowrite8(reg,val)} read and write 8 bits of data from and to a memory-mapped register, respectively.

%MMIO involves mapping an I/O peripheral device into the kernel's address space, \eg via \code{ioremap}~\cite{ioremap}. 
%The mapping creates a one-to-one correspondence between a range in the kernel's virtual address space and a region in physical memory. As such, a read or write instruction on a target virtual address translates to a read/write on the corresponding physical address.
\subpoint{DMA.~}It is a mechanism used by peripheral devices to transfer I/O data to and from main memory bypassing the processor~\cite{corbet2005linux}.
A specialised hardware component, \ie the \textit{DMA controller}, coordinates the DMA data transfer, and generates an interrupt request to the CPU upon DMA completion.
%, after which the CPU reads the transferred data.\jm{not sure whether that last part of sentence is correct. Why "the CPU reads the transferred data"? DMA avoids it.}

\begin{figure}[!t]
	\centering
	\includegraphics[scale=0.75]{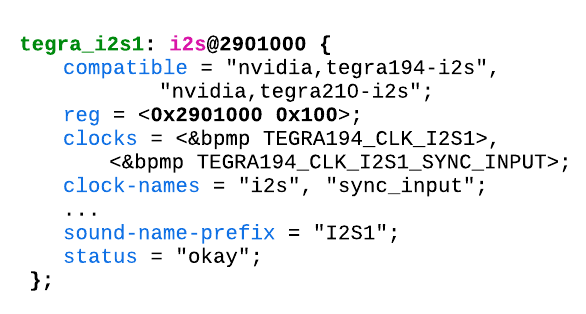}
	\caption{DT node for an I2S interface on Tegra 194 SoC.}
	\label{fig:i2s-dts}
	\vspace{-3mm}
\end{figure}
MMIO or DMA memory for a particular peripheral device is usually associated with a \textit{base address} and \textit{size}, which indicate respectively the base address in physical memory, and the size of the contiguous portion of memory used by the data transfer mechanism. 
The peripheral's specifications provides information regarding base addresses, sizes, and the offsets for all useful registers and DMA regions. 
The details regarding all hardware devices on a system are described in a data structure called the \textit{device tree} (DT). 
The DT contains nodes describing all the hardware on a system, including CPUs, memory, clocks, and peripheral devices like \iis , Ethernet cards, \etc. 
For example, \autoref{fig:i2s-dts} represents a node in the DT of an Nvidia Jetson AGX Xavier kit describing \iis peripheral's properties, such as the base address in memory (\code{0x2901000}) and size (\code{0x100}). The DT is read by the kernel at boot time.
%\vs{add a sentence here to explain why you show the one for the Jetson AGX. Something like ' This specific ARM board is used in our experimental evaluation'}

%\smallskip\noindent\textit{Generic structure of a Linux device driver}
%\py{TODO}

%\subsection{Machine learning classification}
%\py{TODO}

%\subsection{WebAssembly (WASM)}
%\py{TODO: a short intro on wasm since it is used to run the ml model in optee}

%\smallskip\noindent\textbf{Device tree structure}
%\label{sec:dts}

%!TEX root = main.tex
\section{Threat model}
\label{sec:tmodel}
\sys has four main security goals, aimed to guarantee the confidentiality and integrity of sensitive data (\eg images, voice recordings) originating from IoT peripherals before its transmission to the cloud:

\begin{enumerate}[label={(\bfseries G\arabic*):}, leftmargin=28pt]
	\item Protect the confidentiality of peripheral data from the underlying OS or hypervisor.
	\item Ensure a trusted path between kernel space where the data is originally obtained to secure user space where data processing (\eg, encoding, decoding, filtering) occurs.
	\item Ensure the confidentiality (and integrity) of sensitive peripheral data transferred to the cloud, or complete removal of sensitive data from the data stream.%\jm{integrity really? Also, hard to see the difference with G1.}
	\item Minimise the trusted computing base and hence the potential attack surface.
\end{enumerate}

\subpoint{Hardware.~} We assume that the underlying hardware itself is not malicious and cannot be tampered with by the adversary.
\sys shields existing peripherical devices connected to the SoC, and does not consider the connection of new components, \ie the adversary's goal is to access sensitive data of already connected peripherals, typically via MMIO or DMA.
Denial-of-service attacks (\eg maliciously shutting down the system) as well as side-channel vulnerabilities~\cite{liu} are considered out-of-scope.

\subpoint{Software.~} The on-die boot ROM and intermediate firmware, as well as OP-TEE are considered trusted components, which permit establishment of a \textit{chain-of-trust} during a secure boot process.
We assume the peripheral's hardware components information, \eg MMIO or DMA physical addresses, is encapsulated in a device tree file that is signed and verified while establishing the chain-of-trust, preventing a potential attacker from misconfiguring the device.
All other software on the system is considered untrusted.

\subpoint{Third-party.~}
Any element outside the hardware perimeter of the SoC, such as the cloud providers in the IoT setup (\eg AWS IoT~\cite{awsiot}) to which potentially sensitive peripheral data is transmitted, are considered untrusted.

%\py{malicious OS: malware can be installed inside. Secure boot process permits to establish a root of trust which ensures no malicious code presence in secure world.}

%\py{the aim of the adversary is to obtain sensitive information, \eg sound, images. DoS attacks should be out of scope.}

%\smallskip\noindent\textit{Choice of TEE technology.}
%The TEE technologies outlined previously have different security properties and are adapted for different systems and threat models. ARM is the dominant architecture for user-end devices in IoT platforms, making TrustZone technology the best choice when compared to security technologies like Intel SGX, TDX, or AMD SEV which are adapted for server-end environments. As such, we base the design of \sys on ARM TrustZone.
%!TEX root = main.tex

\section{\sys architecture}
\label{sec:arch}
\begin{figure}[!t]
	\centering
	\includegraphics[scale=0.55]{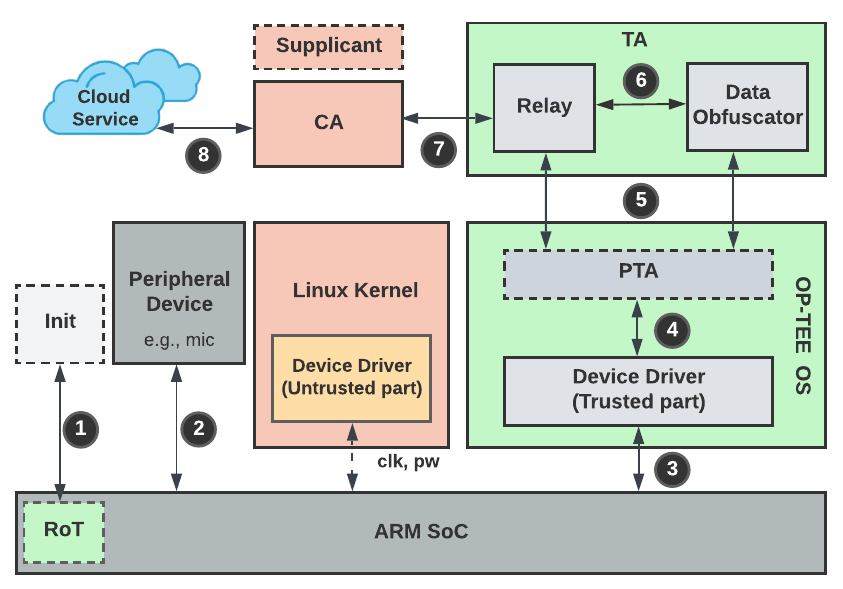}
	\caption{\sys workflow.}
	\label{fig:arch}
	\vspace{-10mm}
\end{figure}
\subsection{Overview}
%The goal of \sys is to shield data generated by a hardware peripheral device, \eg microphone, camera, \etc from an untrusted OS or hypervisor, and obfuscate or filter-out any sensitive information before it can make its way to an untrusted cloud provider. %\sys achieves this goal with a combination of TEE and ML techniques.

In this section, we present the architecture and design of \sys, and show how its components interact to achieve the security goals outlined in \S\ref{sec:tmodel}.

The workflow of \sys is summarised in \autoref{fig:arch}.
At system initialisation, the secure boot process authenticates system boot components, \eg OP-TEE OS image, DT files, and TZASC is used to configure peripheral MMIO and DMA memory regions to be accessible to the secure world only (\ding{202}).
After system setup, a peripheral device, \eg microphone or camera which constitutes a smart home/IoT setup, generates potentially sensitive data, \ie speech or visual data (\ding{203}).
In a regular setup, the device driver software is part of the untrusted OS, thus leaking sensitive data.
In \sys, the driver is partitioned into trusted and untrusted parts, which respectively execute in the secure and normal world.
The trusted part is embedded in the OP-TEE OS kernel, and has exclusive access to the generated peripheral data, which is read into secure I/O buffers (\ding{204}).
The sensitive data is securely processed by the trusted driver, after which it is transferred to a trusted application via the PTA interface (\ding{205}--\ding{206}).
The TA comprises a data obfuscator that acts as a firewall, encrypting or filtering out any sensitive information (\ding{207}).
Converting human speech into a finite set of voice commands is a typical example of stripping potentially sensitive information.
Finally, the reviewed data is sent to an untrusted cloud service via a relay module in the TA (\ding{208}--\ding{209}).
The relay module leverages a Linux user-space daemon called the \textit{TEE supplicant} to provide OS-level services such as network communication or storage.

\begin{figure}[!t]
	\centering
	\includegraphics[scale=0.525]{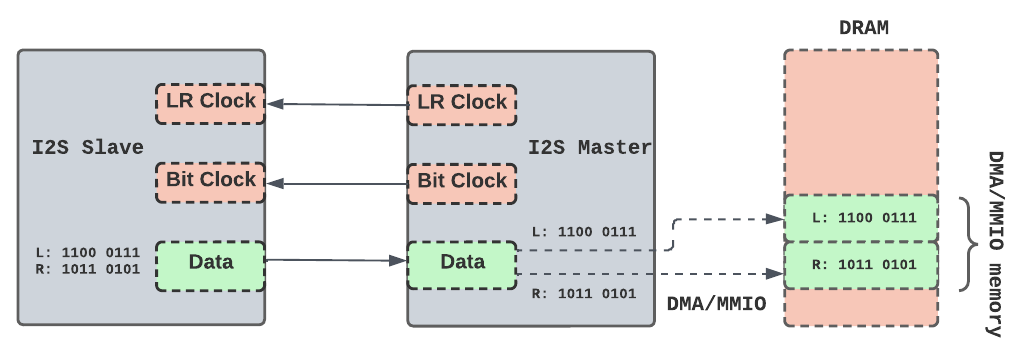}
	\caption{Overview of \iis protocol. One IC, \ie SoC (master) controls data transfer by manipulating two clock lines: the \textit{left-right clock~}~to~specify~the~audio channel (\code{L=0;R=1}), and the \textit{bit clock}~to specify if data can be sent (\code{1}) or not (\code{0}). The second IC, \ie \iis microphone (slave) sends data bits (via MMIO or DMA) for the corresponding channel each time the bit clock is set.}
	\label{fig:i2s-proto}
	%\vspace{-4mm}
\end{figure}
We create a proof-of-concept implementation of this approach with secure sound processing via \iis, a serial communication protocol commonly used to transmit audio data between integrated circuits (ICs). \autoref{fig:i2s-proto} provides an overview of the \iis protocol.

%\py{TODO: explain basics of \iis and how an \iis driver is structured, \ie the main routines/functions involved: clocking, mmio for status/ctrl regs, interrupt for dma, dma rx/tx, power mngt when idle (suspend/resume), init-recording from CA--> driver updates status regs. A short code snippet could be useful. This can be used as a basis to explain how the driver code is partitioned.}

\subsection{System~initialization~and~peripheral memory isolation}

\subpoint{Secure boot.~} To ensure the authenticity and integrity of trusted system components, ARM TrustZone supports \textit{trusted board boot} (TBB)~\cite{tbb}.
The latter establishes a chain-of-trust consisting of several stages of integrity verifications, starting from the boot ROM, comprising the root-of-trust (RoT), and extending through various bootloader stages, the Arm trusted firmware (containing the secure monitor), and OP-TEE OS components.
During the deployment phase, critical components of the TCB, such as \optee OS, device tree files, and TAs undergo cryptographic signing (using SHA-256) with a private key.
The secure boot process uses the corresponding public key to verify the signatures of these components.
By verifying the integrity of \optee OS, this in turn ensures the integrity of \sys's trusted peripheral driver, as well as device tree files containing peripheral I/O registers.

\subpoint{Memory isolation.~} To safeguard peripheral data from unauthorised access by untrusted components like the OS and hypervisor, it is necessary to confine the memory regions associated with peripheral data to the secure world.
In the context of our work, we collectively refer to these memory regions as the \textit{secure I/O region} of the peripheral.
Since most IoT hardware peripherals transfer data using MMIO or DMA mechanisms, \sys configures peripheral MMIO and DMA memory ranges as the secure I/O region.
%\jm{Should we remove this sentence? It seems redundant with the next one.}\py{Redundant why ?}
%For \iis communication, modern SoCs employ either MMIO memory or DMA channels with transmission (TX) and reception (RX) buffers to transmit audio data.
%In \sys, we focus on securing the peripheral's MMIO and DMA memory ranges, which correspond to its secure I/O region.
The specific MMIO registers and DMA regions used by the peripheral are typically defined in the device's DT node.
Our solution offers two methods for specifying these secure I/O regions.
The first is static and involves hard-coding the physical memory addresses of the MMIO and DMA base registers (along with their sizes) directly into the driver's source code.
The second approach is dynamic, where the information is read from the DT node in the corresponding device tree file.

During system initialisation, the trusted firmware utilises TZASC to configure DRAM in such a way that the peripheral's secure I/O region (\ie MMIO and DMA memory ranges) is exclusively accessible to the secure world.
This configuration effectively prevents any access to the peripheral's data from the untrusted OS or hypervisor, thereby satisfying the security requirement in \textbf{G1}.
To complete the initialisation of the peripheral's memory, the trusted driver's initialisation code invokes \optee OS's \code{core\_mmu\_add\_mapping} routine, which maps the secure MMIO regions into kernel-space, ensuring that the secure driver can effectively interact with the peripheral.

%This information is then used by TZASC to configure the secure world's DRAM region to include the peripheral's secure I/O region. As explained previously, TZASC restricts the memory to the secure world only by enforcing memory access checks at runtime.

%DMA control registes are also work via MMIO

%\py{I2S recording command is triggered from normal world. After context switch into TA, recording is triggered via the driver/PTA which sets the required TX/RX registers that allow I2S data to be transmitted correctly.}

%\py{TODO: secure boot, driver/dts file measuring}

%\vspace{-3pt}
\subsection{Secure kernel to user-space communication}
\label{sec:pta-ta-comm}
After the trusted driver reads peripheral data via MMIO or DMA into its I/O buffers, it needs to transfer the data to a secure user-space TA for further processing operations, \eg encoding, encryption, or filtering.
To establish this secure connection between the \optee driver and the TA, we leverage an \optee PTA.
The latter acts as a bridge, offering an interface that connects the TA to secure driver space.
The TA utilises GlobalPlatform Internal API to safely retrieve peripheral data from the driver's memory space via a PTA invocation.
This operation requires a context switch from S-EL0 to S-EL1, and can be summarised in four steps: \emph{(1)} checking access rights of the TA's destination buffers in the invocation, \emph{(2)} copying necessary parameters (\eg the PTA/driver function identifier) from TA memory to \optee OS memory space, \emph{(3)} issuing a system call to \optee OS to invoke the corresponding PTA/driver routine, and \emph{(4)} copying results from \optee OS memory into the TA's destination buffer.
The creation of this secure channel from \optee OS to the TA satisfies \textbf{G2}.

\subsection{Data obfuscation}
\label{sec:data-obfuscation}
\textit{Data obfuscation} is the process of transforming sensitive information to safeguard it from unauthorized access.
Common data obfuscation techniques include encryption, randomization, anonymization \etc.
By integrating a data obfuscation module in \sys, sensitive data originating from an IoT peripheral can be protected before its transmission to an untrusted party in the cloud or the host OS, hence satisfying \textbf{G3}.
The generic architecture proposed by \sys offers the flexibility to incorporate diverse obfuscation techniques depending on the nature of the data.
Our implementation leverages techniques provided by OP-TEE's cryptography API which is based on LibTomCrypt~\cite{libtomcrypt}, a popular open-source cryptographic library.
The latter provides various symmetric cryptographic algorithms, including AES-GCM, which can be leveraged to encrypt and guarantee the integrity of sensitive data. 
%Public-key encryption techniques, although viable, were not selected for our design due to their performance overhead when compared to their symmetric counterpart.
Additional data obfuscation methods, including \emph{data conversion} and \emph{data filtering}, serve to reduce the sensitivity of the data or completely strip sensitive elements, respectively.
Data conversion can be particularly effective for human voice recordings, converting them into voice commands upon recognition of specific voice patterns.
Data filtering, on the other hand, can entirely remove unrecognised sentences.
The primary advantage of data obfuscation is that it effectively restricts the distribution of sensitive data in uncontrolled environments, such as the REE and the cloud service providers.

\subsection{Driver partitioning}
\label{sec:partitioning}
\begin{figure}	
	\centering
	\includegraphics[scale=0.525]{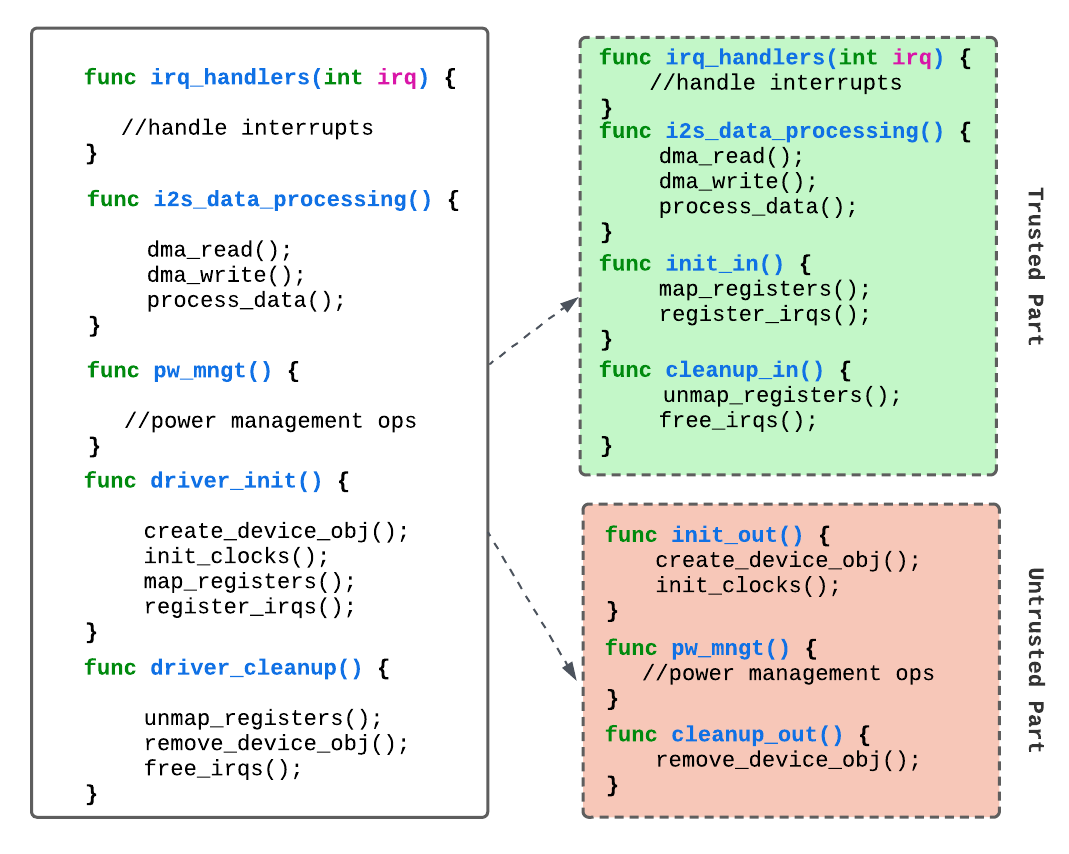}
	\caption{\sys driver partitioning.}
	\vspace{-4mm}
	\label{fig:driver-part}
\end{figure}
A key aspect of TEE development is decreasing the \textit{trusted computing base} (TCB). The latter represents the components of a system that must be trusted to enforce the system's security policies. Decreasing the TCB reduces the potential attack surface and the likelihood of vulnerabilities in the TEE. TCB reduction is usually done via \textit{code partitioning}. This involves identifying portions of the code that manipulate sensitive data and isolating only these within the TEE; the non-sensitive part can be kept out of the TEE. OP-TEE provides some examples of how to build secure peripheral drivers but lacks a systematic approach for partitioning them into trusted and untrusted components.

This section outlines a high-level approach to assist developers in partitioning drivers for securing peripheral data in \sys.
Using a simple \iis peripheral driver as an example, we manually partition its code into trusted and untrusted parts.
The trusted part includes the code which interacts with the peripheral's secure I/O region.
For \iis, the trusted driver handles tasks like mapping MMIO registers, read/write operations, and DMA buffer management.
The untrusted part deals with non-sensitive operations like clock and power-management.
\autoref{fig:driver-part} provides a generic blueprint for manual partitioning in \sys, satisfying \textbf{G4}. 
Advanced techniques like static data-analysis may be used for more complex code bases to pinpoint code portions interacting with sensitive regions of the peripheral.

%The aim of this section is to provide a high-level approach to guide developers in partitioning drivers for securing peripheral data in \sys. We illustrate our approach by considering a simple \iis peripheral driver to be partitioned.  
%
%
%In \sys, we perform \textit{manual partitioning} of the driver into \textit{trusted} and \textit{untrusted} parts.
%To determine which part of the driver to secure in the TEE, one must first identify the driver code that interacts with the peripheral's secure I/O region.
%In the case of \iis, the microphone generates audio data which is transferred to memory via DMA or MMIO.
%MMIO registers are provided by the SoC used for \iis control, such as enabling/disabling \iis, or specifying audio data formats.
%In this case, the trusted driver code integrates the complete logic required for mapping MMIO registers, for performing read and write operations on these MMIO registers and DMA buffers.
%Additionally, the code manages interrupt handling, such as when a DMA operation terminates.
%The remaining code, which generally comprises clock and power management operations, is delegated to the untrusted part driver since this code doesn't access the data being secured.
%\autoref{fig:driver-part} provides a generic blueprint for manually partitioning a driver with \sys, satisfying \textbf{G4}.
%For more complex code bases, techniques such as \textit{static data-analysis}~\cite{aces,minionkim2018securing} could be leveraged to properly identify which code portions interact with sensitive MMIO or DMA regions of the peripheral. 

%!TEX root = main.tex
\section{Implementation}
\label{sec:implem}

\subpoint{Communication between TA and PTA.~}
A user-space TA uses the GlobalPlatform TEE API function \code{TEE\_OpenTASession} to initialise a session with the \iis PTA using the latter's UUID.
After the session is created, the TA uses the \code{TEE\_InvokeTACommand} with a command identifier and related parameters stored in the structure \code{TEE\_Param} to invoke a service, \eg initialising an \iis recording operation in the PTA.
The PTA in turn invokes the corresponding secure \iis driver routine, which performs secure I/O operations via MMIO or DMA.
The resulting data is transferred securely to the TA in a \code{TEE\_Param memref} buffer.
The data is then encrypted (or filtered) inside the TA prior to transmission to the cloud via the TEE supplicant executing inside the REE.

\subpoint{Key derivation and storage.~}
\optee features a built-in PTA specifically designed for the derivation of encryption keys based on hardware-unique identifiers, exposed by SoCs that offer such capabilities.
These hardware-specific keys are accessible exclusively in the TEE, or are derived differently by the SoC depending on whether they are accessed from the REE or TEE.
Within the TEE, the resulting encryption keys are commonly used for encryption and decryption of files, ensuring data security even after device reboots.
This functionality in OP-TEE is known as \textit{secure storage}~\cite{optee-sec-storage}.
Other research prototypes serve such uniquely-derived cryptographic materials as the basis for introducing attestation capabilities into TrustZone, fulfilling a gap in Arm's TEE architecture.
This feature is essential for establishing the trustworthiness of executing software, thus offering a fundamental security assurance for third-party actors, including cloud service providers~\cite{DBLP:conf/dais/MenetreyGKPFSR22,DBLP:conf/icdcs/MenetreyPFS22}.

%!TEX root = main.tex
\section{Evaluation}
\label{sec:eval}
This section presents an experimental evaluation of \sys. We seek to answer the following questions: 

\begin{itemize}[]
	\item[$Q_1$)] What is the overhead of reading and writing MMIO registers in a TEE? (\S\ref{sec:eval-mmio})
	%\item[$Q_2$)] What is the overhead of doing driver operations inside a TEE?
	%(\S\ref{sec:eval-i2s-recording})
	\item[$Q_2$)] What is cost of data transfers between the secure driver and a trusted application? (\S\ref{sec:eval-copy})
	%\item[$Q_4$)] What is the impact of performing machine learning classification in a TA? (\S\ref{sec:eval-ml-classifier})
		\item[$Q_3$)] What is the cost of data obfuscation inside a TA? (\S\ref{sec:eval-crypto})
	\item[$Q_4$)] What is the degree of TCB reduction after driver partitioning? (\S\ref{sec:eval-tcb-reduc})
	%\item System energy usage 

\end{itemize}

\smallskip\noindent\textit{Experimental setup.} 
Our evaluations were conducted on an Nvidia Jetson AGX Xavier development kit.
This kit ships with 8 TrustZone-enabled ARMv8.2 (64 bit) processors (4 active) each clocked at 2265 MHz, and 32 GB of memory.
The kit runs Jetson Linux 35.3.1, which is based on Ubuntu 20.04.5 LTS and Linux Kernel 5.10. The OP-TEE OS version used is based on tag \code{optee\_os-nvidia-r35.2.1} of NVIDIA OP-TEE OS fork~\cite{nvidia-optee-os}. %\py{CPU freq: min = 115.2000; max = 2265.6001 Hz}
Latencies (in clock cycles) are measured by reading the \code{CNTVCT\_EL0} timer register~\cite{cntvct}.
The frequency of this timer is 31.25 MHz on the experimental system.
This timer cannot be accessed directly by a TA; we access its value from TAs via an \optee PTA interface.
Unless stated otherwise, we report the average of 100 runs.

\subsection{Performance of MMIO operations}
\label{sec:eval-mmio}

\begin{figure}
	%\vspace{-4mm}
	\centering
	\includegraphics[scale=0.775]{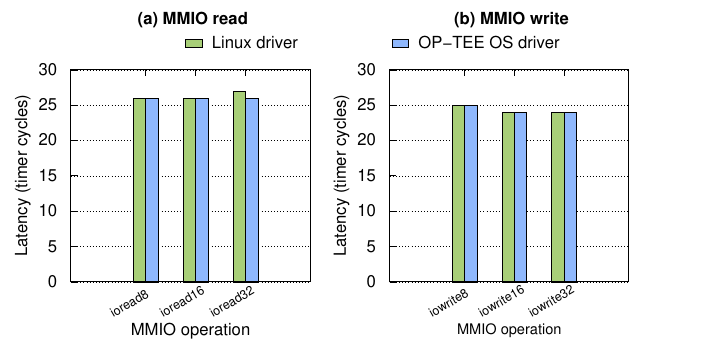}
	\caption{Cost of MMIO operations.}
	\label{fig:mmio-read-write}
\end{figure}
%In our initial experiment, we assess the performance of MMIO operations on an MMIO register in both the secure and normal worlds, using different MMIO APIs, \ie \code{ioread\{8,16,32\}}, \code{iowrite\{8,16,32\}}. Our findings, as depicted in \autoref{fig:mmio-read-write}, indicate that MMIO read and write operations exhibit identical performance when executed in a standard normal-world Linux Kernel driver and a secure OPTEE OS driver. This similarity in performance can be attributed to the fact that MMIO operations share the same underlying implementations in both secure and normal worlds, and are subject to uniform access control policies (\ie TZASC checks) across both worlds. The reads and writes have identical performance for 8,16, and 32 bits, $\approx 26$ cycles for reads, and $\approx 24$ cycles for writes. This is because the MMIO register width is the same (32-bit) on the device. This means the full 32 bits are read or written, and truncated accordingly to align with the read or write size.xx
(\emph{Answer to Q1})~
In our initial experiment, we assess the performance of MMIO operations in both secure and normal worlds, using different MMIO APIs, \ie \code{ioread\{8,16,32\}}, \code{iowrite\{8,16,32\}}.
 Our findings, as depicted in \autoref{fig:mmio-read-write}, indicate that MMIO read operations in a secure \optee driver exhibit identical performance as those in a standard normal-world Linux kernel driver.
 The same observation is true for MMIO write operations.
 This similarity in performance can be attributed to the fact that MMIO operations share the same underlying implementations in both secure and normal worlds, and are subject to uniform access control policies (\ie TZASC checks) across both worlds.
 The read operations have similar performance ($\approx 26$ cycles) for 8,16, and 32 bits.
 We also have a similar cost  with the write operations ($\approx 24$ cycles) for 8,16, and 32 bits.
 This is because the MMIO register width is the same (32-bit) on the device, meaning the full 32 bits are always read or written, and truncated accordingly to align with the exact read or write size.
\observ{Kernel driver MMIO operations can be safeguarded within \optee without sacrificing performance.}

%\subsection{Overhead of secure driver vs non-secure one}
%\label{sec:eval-dma}
%\py{TODO: polling the TX/RX channels and reading audio data for Xs}

%\observ{overall cost of driver operations is similar in TEE}

\subsection{Cost of buffer copying operations}
\label{sec:eval-copy}

\begin{figure}
	\includegraphics[scale=0.775]{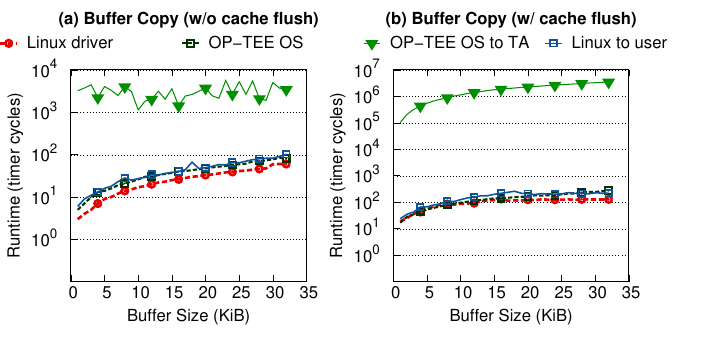}
	\caption{Cost of buffer copying operations.}
	\label{fig:buffer-copy}
	\vspace{-3.5mm}
\end{figure}
(\emph{Answer to Q2})~
This experiment aims to evaluate the cost of transferring data from secure kernel space (\optee OS) to secure user space (TA).
It involves the TA invoking an \optee OS driver function through the PTA interface using the function: \code{TEE\_InvokeTACommand}.
The TA passes a parameter buffer of type: \code{TEE\_Param} as input, and data is copied (using \code{memcpy}) from a buffer of the secure driver into the destination buffer.
We compare this operation against buffer copy operations from normal world kernel space (Linux driver) to normal world user space, using the \code{copy\_to\_user} primitive provided by the Linux kernel. We also evaluate the cost of buffer copy operations in normal and secure world kernel space only, using standard \code{memcpy} operations.
The copy operations are done with and without cache flushing enabled for the source and destination buffers.
\autoref{fig:buffer-copy} outlines the results, which indicate that copying a buffer from the secure driver in \optee OS to the TA is about $113\times$ and $10K\times$ more expensive on average when compared to copying a buffer from a normal world Linux kernel buffer to a normal world user space buffer, without and with cache flushing respectively.
%$128\times$ and $12K\times$ more expensive on average when compared to standard \code{memcpy} operations with and without cache flushes respectively within \optee OS.
%rom the TA is $\approx 630$ cycles (300-700 cycles).A round trip from the TA to the PTA and back is $\approx 89620$ cycles on our experimental system.
The high overhead observed in the copy operation from \optee OS to the TA is attributed to the additional memory access checks and buffer copying operations involved when transferring data between the TA and \optee OS, as outlined in \S\ref{sec:pta-ta-comm}.

\observ{Copying a buffer from \optee OS to a TA is expensive due to additional copy operations and security checks performed during the S-EL0 to S-EL1 context switch.}
%\subsection{Cost of DMA transfers in a TEE vs a non-TEE environment}
%\label{sec:eval-dma}
%\py{TODO: polling the TX/RX channels and reading audio data for Xs}

\subsection{Cost~of~data~encryption/decryption operations}
\label{sec:eval-crypto}
\begin{figure}
	%\vspace{-3mm}
	\centering
	\includegraphics[scale=0.75]{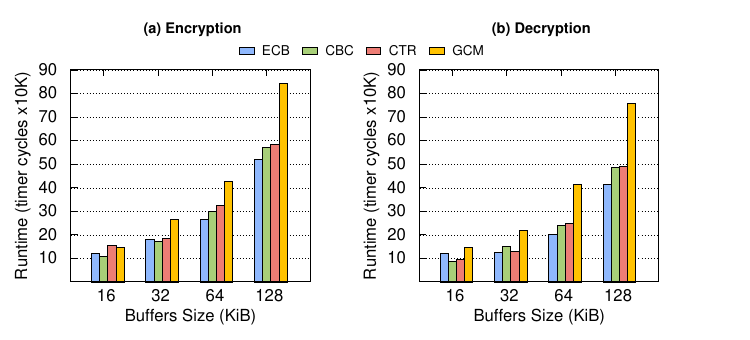}
	\caption{Cost of cryptographic operations inside a TA.}
	\label{fig:crypto}
	\vspace{-2mm}
\end{figure}
(\emph{Answer to Q3})~
In this experiment we evaluate the encryption and decryption costs inside a TA of four AES cryptographic modes of operation: ECB, CBC, CTR, and GCM.
All the cryptographic operations use the same 128-bit key. AES-GCM uses an initialisation vector and tag length of 128 bits.
\autoref{fig:crypto} shows the results obtained.

%\py{NB: larger buffer were used here because the time does not vary interestingly from 0-32}

ECB is relatively simple and deterministic, \ie identical plaintext blocks produce identical ciphertext blocks; while this mode is relatively cheaper in cost, it is not recommended for securing sensitive data.
 More advanced algorithms like CBC introduce block chaining, which improves security.
 CTR mode has similar performance and security as compared to CBC.
 GCM mode is most expensive in general because it computes an authentication tag to ensure data integrity, resulting in larger computational overhead.
 On average, GCM is about $1.3\times$ and $1.6\times$ slower compared to CTR mode for encryption and decryption, respectively.
 However, in spite of the relatively higher cost, GCM provides both confidentiality and integrity guarantees, while the other modes only ensure confidentility.

\observ{Encryption techniques like AES-GCM can be leveraged in the TA to provide confidentiality and integrity guarantees to peripheral data.}

\subsection{TCB analysis}
\label{sec:eval-tcb-reduc}
\begin{table}[!b]
	\caption{Source code \% of \iis driver components in the trusted and untrusted drivers after partitioning.}
	%\vspace{-3mm}
	%\centering
	\setlength{\tabcolsep}{5pt}
	\rowcolors{1}{gray!0}{gray!10}
	\resizebox{\columnwidth}{!}{
		\begin{tabularx}{\linewidth}{lcc}
			\toprule			
			\textbf{Component} & \textbf{\% trusted driver} & \textbf{\% untrusted driver}\\
			\midrule
			
			Initialization & 22.13 & 23.45 \\
			Data processing & 41.15 & 0\\
			IRQ handling & 1.33 & 0\\
			Power management & 0 & 7.52\\
			Cleanup & 2.21 & 2.21\\
			\midrule
			\textbf{Total} & \textbf{66.82} & \textbf{33.18}\\
			\bottomrule
		\end{tabularx}
	}
	%\vspace{2mm}	
	\label{tab:tcb}
\end{table}
(\emph{Answer to Q4})~
\autoref{tab:tcb} provides a summary of the code components comprising the partitioned \iis driver, along with the proportion of source code they represent (wrt. the unpartitioned version) in terms of lines of code (LOC).
Following the partitioning approach outlined in \S\ref{sec:partitioning}, $66.82\%$ of the source code was secured inside \optee OS's trusted driver, which corresponds to a $33.18\%$ decrease in the TCB with respect to a system which includes all the source code inside the secure \optee OS driver. 

\observ{A decrease in the TCB holds significant security implications because it reduces the attack surface available to potential adversaries, and mitigates the effects of security vulnerabilities that may exist in the code.}

%The power management operations comprise of instructions for gating and ungating \iis related registers.
%\py{TODO: tcb reduction \%, discussion:security analysis, threat mitigation}

%!TEX root = main.tex
\section{Related work}
\label{sec:rw}

\begin{table}[t]
\caption{Comparing \sys with previous approaches.}
%\vspace{-3mm}
%\centering
\setlength{\tabcolsep}{1.5pt}
\rowcolors{1}{gray!0}{gray!10}
\resizebox{\columnwidth}{!}{
\begin{tabularx}{\linewidth}{Xcccccc}
	%\toprule			
	& \rotatebox{0}{\textbf{~\cite{aces}}} & \rotatebox{0}{\textbf{~\cite{driverguard}}} & \rotatebox{0}{\textbf{~\cite{secloak}}} & \textbf{~\cite{teep}} & \textbf{~\cite{sgxio}} & \rotatebox{0}{\textbf{Us}} \\
	\midrule
	Untrusted OS              & \no   & \yes & \no & \yes & \yes & \yes \\
	Untrusted hypervisor      & \no   & \no & \no & \no & \no & \yes \\
	Data confidentiality      & \no   & \yes & \no & \no & \yes & \yes \\
	Data integrity            & \yes  & \yes & \yes & \no & \yes & \yes \\
	ARM-based						  & \yes   & \no & \yes & \yes & \no & \yes \\
	\bottomrule
\end{tabularx}
}
\vspace{-4mm}
\label{tab:rw}
\end{table}

Several research works have proposed techniques to enhance security guarantees of peripheral data both on ARM and x86-based systems.
For ARM-based systems: \cite{liu2012software} provides software abstractions to protect the integrity and authenticity of sensor readings; SeCloak~\cite{secloak} leverages TrustZone to provide a way to reliably turn on/off peripherals even in the presence of compromised privileged platform software like the OS or hypervisor, without explicitly providing data confidentiality or integrity guarantees; other systems like~\cite{aces} leverage an MPU to isolate peripheral memory in MCUs.
For x86-based systems: SGXIO~\cite{sgxio} and Driverguard~\cite{driverguard} leverage a trusted hypervisor to secure sensitive I/O data from an untrusted OS.
While these systems propose varied solutions to address I/O data security, they either weaken the threat model by leveraging a trusted OS or hypervisor, or do not provide full confidentiality and integrity guarantees.
As summarized in \autoref{tab:rw}, \sys addresses all these limitations and offers a completely secure path for IoT peripheral data from the hardware to the cloud, safeguarding data confidentiality and integrity. 

%\sys on the other hand safeguards the confidentiality and integrity of IoT peripheral data from malicious privileged software and untrusted clouds.

%Safe and Practical GPU Computation in TrustZone~\cite{gpuTz}\py{TODO}
%Heterogeneous Isolated Execution for Commodity GPUs~\cite{hix}

%The Design and Implementation of Microdrivers~\cite{microdrivers}\py{ASPLOS'2008: partitions drivers into user and kernel parts.}

%Confidential Execution of Deep Learning Inference at the Untrusted Edge with ARM TrustZone~\cite{tz-dlinf}
%Protecting Data on Smartphones and Tablets from Memory Attacks~\cite{colp15}

%TEEp: Supporting Secure Parallel Processing in ARM TrustZone~\cite{teep}
%see: \url{https://stefan.t8k2.com/publications/mobisys/2012/tenor.pdf}

%!TEX root = main.tex
\section{Practical applications of \sys}
\label{sec:discussion}
The security architecture provided by \sys can be leveraged in various IoT contexts:

\subpoint{Smart homes.~}
In a smart home setup, the central hub can leverage \sys framework to enhance data privacy.
\sys acts as a safeguard, stopping the unintentional sharing of sensitive information like surveillance videos or voice recordings to cloud services with data obfuscation.
For example, as explained in \S\ref{sec:data-obfuscation}, \sys can apply data conversion to human speech, converting it into voice commands.
This avoids the need to send raw audio (containing potentially sensitive information) for analysis; only the specific voice command can be identified and sent to the cloud service.
Implementing such strategies requires moving machine learning-based inference into the TEE, a paradigm that can be achieved with Arm TrustZone~\cite{DBLP:conf/mobisys/MoSKDLCH20}.
Additionally, peripherals like fingerprint sensors employed for locking mechanisms can be effectively isolated from potential malicious attacks that compromise the OS.
The sensor reading, as well as all necessary logic to validate the latter is protected within the TEE.
%\jm{For me, this is unclear how what attack is prevented, and how \sys solves it. We either need to expand, or dismiss it.}

\subpoint{Medical IoT.~}
Similarly to a smart home, medical IoT hubs responsible for aggregating potentially sensitive data from devices such as cardiac monitors can leverage \sys to guarantee that this data is not accessible to a malicious adversary (\eg on the LAN) who has compromised the OS. Further, the data obfuscation process safeguards the sensitive data once it is transferred to an untrusted cloud service.

%In IoT setups, these recording operations can be periodic, \eg in the context of home surveillance, gathering sensor readings, \etc.
%!TEX root = main.tex
\section{Limitations and discussions}
\label{sec:limitations}
Manual partitioning can be very challenging and error-prone for large code bases, as there may be complex data paths with indirect access to peripheral data, which are difficult to identify manually.
Nonetheless, in \sys, restricting the peripheral's secure I/O region to the secure world always prevents access to peripheral data from untrusted code, even if the latter contains instructions that manipulate peripheral data.
To overcome the complexities inherent in manual partitioning, techniques such as static data-analysis~\cite{aces,minionkim2018securing} could be leveraged to properly identify all code parts that interact (directly or indirectly) with sensitive data.

%Nothwithstanding tis statically hard-coded inside the codehe extensive code coverage provided by static analysis, the latter is not sufficient for partitioning code and must be 

%\smallskip\noindent\textit{Performance.} IoT devices generally have limited processing power, making it challenging to run complex ML models. Further, TEE technologies like TrustZone provide relatively small memory resources for applications, which is inconvenient for large ML models.
%!TEX root = main.tex
\section{Conclusion}
\label{sec:conclusion}
%This paper presents \sys, a generic design to secure sensitive data in IoT devices by leveraging trusted execution environments and machine learning. \sys shields peripheral driver code in a TEE, and leverages ML classification to filter out sensitive information before it can make its way to an untrusted cloud. We build a proof-of-concept implementation with \iis and our evaluation results show \sys improves security at a reasonable cost. \py{future work: adding machine learning to process data}

This paper presents \sys, a generic design to secure sensitive peripheral data in IoT environments by leveraging TEE technology.
\sys restricts peripheral I/O memory regions to a secure kernel-space TEE, thereby limiting access exclusively to an small segment of the peripheral driver code.
This strategy effectively blocks unauthorised access from potentially compromised operating systems or hypervisors.
The data is then securely transferred to a user-space TEE, where obfuscation techniques are applied prior to its transmission to an untrusted host operating system or cloud service providers.
Through a proof-of-concept implementation focusing on \iis, our evaluations show that \sys achieves enhanced the security posture of IoT devices at a reasonable cost.

\smallskip\noindent\textit{Future work}.~
We aim to extend our work along two axes: firstly by integrating static analysis techniques to streamline the code partitioning process, and subsequently by introducing machine learning classification algorithms to automate data obfuscation.
The latter approach involves the use of pre-trained deep learning models within the TA to identify sensitive data in data streams, upon which appropriate obfuscation techniques may be applied.

\section*{Acknowledgments}
This work was supported by the VEDLIoT: Very Efficient Deep Learning in IoT European project (no. 957197) and the Swiss National Science Foundation under project P4: Practical Privacy Preserving Processing (no. 200020\_215216).

\bibliographystyle{plain}
\bibliography{min}

%\appendix

\end{document}